\documentstyle[aps,twocolumn,psfig,subfigure]{revtex}

\begin{document}
\draft
\twocolumn[
\hsize\textwidth\columnwidth\hsize\csname
@twocolumnfalse\endcsname

\title{Many-body approach to the dynamics of batch learning}
\author{K. Y.~Michael Wong, S.~Li, and Y. W.~Tong}
\address{
Department of Physics, Hong Kong University of Science and Technology,
Clear Water Bay, Kowloon, Hong Kong
}
\date{September 1, 1999}
\maketitle


\begin{abstract}
Using the cavity method and diagrammatic methods, 
we model the dynamics of batch learning
of restricted sets of examples,
widely applicable to general learning cost functions,
and fully taking into account the temporal correlations
introduced by the recycling of the examples.
\end{abstract}
]


The extraction of input-output maps from a set of examples, 
usually termed {\it learning}, 
is an important and interesting problem in information processing tasks 
such as classification and regression \cite{bishop}. 
During learning, one defines an energy function 
in terms of a training set of examples, 
which is then minimized by a gradient descent process 
with respect to the parameters defining the input-output map. 
In batch learning, the same restricted set of examples 
is provided for each learning step. 
There have been attempts using statistical physics 
to describe the dynamics of learning with macroscopic variables. 
The major difficulty is that the recycling of the examples 
introduces temporal correlations of the parameters in the learning history. 
Hence previous success has been limited to 
Adaline learning \cite{hertz,opper,krogh}, 
linear perceptrons learning nonlinear rules \cite{bos,bos2},
Hebbian learning \cite{coolen,rae} and binary weights \cite{horner}.

Recent advances in {\it on-line} learning 
are based on the circumvention of this difficulty. 
In contrast to batch learning, an independent example is generated 
for each learning step \cite{saadsolla,saadrattray}. 
Since statistical correlations among the examples can be ignored, 
the dynamics can be simply described by instantaneous dynamical variables.
However, on-line learning represents an ideal case 
in which one has access to an almost infinite training set, 
whereas in many applications, 
the collection of training examples may be costly. 

In this paper, we model batch learning of restricted sets of examples, 
by considering the learning model as a many-body system. 
Each example makes a small contribution to the learning process, 
which can be described by linear response terms 
in a sea of background examples. 
Our theory is widely applicable to any gradient-descent learning rule 
which minimizes an {\it arbitrary} cost function 
in terms of the activation. 
It fully takes into account the temporal correlations during learning, 
and is exact for large networks. 
Preliminary work has been presented recently \cite{wlt}.


Consider the single layer perceptron
with $N\gg 1$ input nodes $\{\xi_j\}$ 
connecting to a single output node by the weights $\{J_j\}$ 
and often, the bias $\theta$ as well. 
For convenience we assume 
that the inputs $\xi_j$ are Gaussian variables with mean 0 and variance 1, 
and the output state $S$ is a function $f(x)$ 
of the {\it activation} $x$ at the output node, i.e.
$S = f(x)$; $x = \vec J\cdot\vec\xi+\theta$. 
For binary outputs, $f(x)={\rm sgn}x$.

The network is assigned to ``learn'' $p\equiv\alpha N$ examples 
which map inputs $\{\xi^\mu_j\}$ to the outputs 
$\{S_\mu\}\ (\mu=1,\dots, p)$. 
In the case of random examples, 
$S_\mu$ are random binary variables, 
and the perceptron is used as a storage device. 
In the case of teacher-generated examples, 
$S_\mu$ are the outputs generated by a teacher perceptron 
with weights $\{B_j\}$ and often, a bias $\phi$ as well, 
namely $S_\mu = f(y_\mu)$;
$y_\mu = \vec B\cdot\vec\xi^\mu+\phi$.

Batch learning is achieved by adjusting the weights $\{J_j\}$ iteratively
so that a certain cost function 
in terms of the activations $\{x_\mu\}$ and the output $S_\mu$
of all examples is minimized.
Hence we consider a general cost function $E = -\sum_\mu g(x_\mu,y_\mu)$.
The precise functional form of $g(x,y)$ depends on 
the adopted learning algorithm. 
In previous studies, 
$g(x,y)=-(S-x)^2/2$ with $S={\rm sgn}y$ 
in Adaline learning \cite{hertz,opper,krogh}, 
and $g(x,y)=xS$ in Hebbian learning \cite{coolen,rae}. 

To ensure that the perceptron is regularized after learning, 
it is customary to introduce a weight decay term. 
In the presence of noise, the gradient descent dynamics 
of the weights is given by
\begin{equation}
	{dJ_j(t)\over dt}
	={1\over N}\sum_\mu g'(x_\mu(t),y_\mu)\xi^\mu_j
	-\lambda J_j(t)+\eta_j(t),
\label{original}
\end{equation}
where the prime represents partial differentiation with respect to $x$, 
$\lambda$ is the weight decay strength, 
and $\eta_j(t)$ is the noise term at temperature $T$ with 
$\langle\eta_j(t)\rangle=0$ and	
$\langle\eta_j(t)\eta_k(s)\rangle=2T\delta_{jk}\delta(t-s)/N$.
The dynamics of the bias $\theta$ is similar, 
except that no bias decay should be present 
according to consistency arguments \cite{bishop},
\begin{equation}
	{d\theta(t)\over dt}
	={1\over N}\sum_\mu g'(x_\mu(t),y_\mu)+\eta_\theta(t).
\label{bias}
\end{equation}


Our theory is the dynamical version of the cavity method \cite{mpv,epl,nips}. 
It uses a self-consistency argument 
to consider what happens when a new example is added to a training set. 
The central quantity in this method is the {\it cavity activation}, 
which is the activation of a new example 
for a perceptron trained without that example. 
Since the original network has no information about the new example, 
the cavity activation is random.
Here we present the theory for $\theta=\phi=0$,
skipping extensions to biased perceptrons.
Denoting the new example by the label 0,
its cavity activation at time $t$ is
$h_0(t)=\vec J(t)\cdot\vec\xi^0$.
For large $N$, $h_0(t)$ is a Gaussian variable. 
Its covariance is given by the correlation function $C(t,s)$ 
of the weights at times $t$ and $s$, that is,
$\langle h_0(t)h_0(s)\rangle
=\vec J(t)\cdot\vec J(s)\equiv C(t,s)$,
where $\xi^0_j$ and $\xi^0_k$ 
are assumed to be independent for $j\ne k$. 
For teacher-generated examples, the distribution is further specified by 
the teacher-student correlation $R(t)$, given by
$\langle h_0(t)y_0\rangle=\vec J(t)\cdot\vec B\equiv R(t)$.

Now suppose the perceptron incorporates the new example 
at the batch-mode learning step at time $s$. 
Then the activation of this new example at a subsequent time $t>s$ 
will no longer be a random variable. 
Furthermore, the activations of the original $p$ examples at time $t$ 
will also be adjusted from $\{x_\mu(t)\}$ to $\{x^0_\mu(t)\}$ 
because of the newcomer, 
which will in turn affect the evolution of the activation of example 0, 
giving rise to the so-called Onsager reaction effects. 
This makes the dynamics complex, 
but fortunately for large $p\sim N$, 
we can assume that the adjustment 
from $x_\mu(t)$ to $x^0_\mu(t)$ is small, 
and linear response theory can be applied.

Suppose the weights of the original and new perceptron at time $t$ 
are $\{J_j(t)\}$ and $\{J^0_j(t)\}$ respectively. 
Then a perturbation of (\ref{original}) yields
\begin{eqnarray}
	&&\left({d\over dt}+\lambda\right)(J^0_j(t)-J_j(t))
	={1\over N}g'(x_0(t),y_0)\xi^0_j
	\nonumber\\
	&&+{1\over N}\sum_{\mu k} \xi^\mu_j g''(x_\mu(t),y_\mu)
	\xi^\mu_k(J^0_k(t)-J_k(t)).
\label{dyneqn}
\end{eqnarray}
The first term on the right hand side describes the primary effects 
of adding example 0 to the training set, 
and is the driving term for the difference between the two perceptrons. 
The second term describes the many-body reactions 
due to the changes of the original examples caused by the added example.
The equation can be solved by the Green's function technique, yielding 
\begin{equation}
	J^0_j(t)-J_j(t)
	=\sum_k\int ds G_{jk}(t,s)
	\left({1\over N}g'_0(s)\xi^0_k\right),
\label{dressed}
\end{equation}
where $g'_0(s)=g'(x_0(s),y_0)$ 
and $G_{jk}(t,s)$ is the {\it weight Green's function}, 
which describes how the effects of a perturbation 
propagates from weight $J_k$ at learning time $s$ to
weight $J_j$ at a subsequent time $t$.
In the present context, the perturbation comes from 
the gradient term of example 0, 
such that integrating over the history and summing over all nodes 
give the resultant change from $J_j(t)$ to $J^0_j(t)$.

For large $N$ the weight Green's function can be found 
by the diagrammatic approach. 
The result is self-averaging 
over the distribution of examples and is diagonal, 
i.e. $\lim_{N\to\infty}G_{jk}(t,s)=G(t,s)\delta_{jk}$, where
\begin{eqnarray}
	G(t,s)
	=&&G^{(0)}(t-s)
	+\alpha\int dt_1\int dt_2 G^{(0)}(t-t_1)\nonumber\\
	&&\langle g''_\mu(t_1)D_\mu(t_1,t_2)\rangle
	G(t_2,s).
\label{wgreen}
\end{eqnarray}
$G^{(0)}(t-s)\equiv\Theta(t-s)\exp(-\lambda(t-s))$
is the bare Green's function, 
and $\Theta$ is the step function.
$D_\mu(t,s)$ is the {\it example Green's function} given by
\begin{equation}
	D_\mu(t,s)
	=\delta(t-s)
        +\int dt'G(t,t')g''_\mu(t')D_\mu(t',s).
\label{fgreen}
\end{equation}
Our approach to the macroscopic description of the learning dynamics 
is to relate the activation of the examples to their cavity counterparts. 
Multiplying both sides of (\ref{dressed}) and summing over $j$, we get
\begin{equation}
	x_0(t)-h_0(t)
        =\int ds G(t,s)g'_0(s).
\label{generic}
\end{equation}
The activation distribution is thus related 
to the cavity activation distribution, 
which is known to be Gaussian. 
In turn, the covariance of this Gaussian distribution 
is provided by the fluctuation-response relation
\begin{eqnarray}
	C(t,s)
        =&&\alpha\int dt'G^{(0)}(t-t')\langle g'_\mu(t')x_\mu(s)\rangle
	\nonumber\\
	&&+2T\int dt'G^{(0)}(t-t')G(s,t').
\label{correlation}
\end{eqnarray}
Furthermore, for teacher-generated examples, 
its mean is related to the teacher-student correlation given by
\begin{equation}
	R(t)
	=\alpha\int dt'G^{(0)}(t-t')\langle g'_\mu(t')y_\mu\rangle.
\label{tscorrelation}
\end{equation}

To monitor the progress of learning, 
we are interested in three performance measures: 
(a) {\it Training error} $\epsilon_t$, 
which is the probability of error for the training examples. 
(b) {\it Test error} $\epsilon_{test}$,  
which is the probability of error 
when the inputs $\xi_j^\mu$ of the training examples 
are corrupted by an additive Gaussian noise of variance $\Delta^2$. 
This is a relevant performance measure 
when the perceptron is applied to process data 
which are the corrupted versions of the training data. 
When $\Delta^2=0$, the test error reduces to the training error. 
(c) {\it Generalization error} $\epsilon_g$ for teacher-generated examples,  
which is the probability of error for an arbitrary input $\xi_j$ 
when the teacher and student outputs are compared. 

The cavity method can be applied to the dynamics of learning
with an arbitrary cost function.
When it is applied to the Hebb rule,
it yields results identical to \cite{coolen}. 
Here for illustration, we present the results for the Adaline rule. 
This is a common learning rule and bears resemblance
with the more common back-propagation rule.
Theoretically, its dynamics is particularly convenient for analysis
since $g''(x)=-1$,
rendering the weight Green's function time translation invariant, 
i.e. $G(t,s)=G(t-s)$.
In this case, the dynamics can be solved by Laplace transform, 
and the cavity approach facilitates a deeper understanding 
than previous studies. 
Illustrative results are summarized with respect to the following aspects:


{\it 1) Overtraining of $\epsilon_g$:} 
As shown in Fig. 1,
$\epsilon_g$ decreases at the initial stage of learning.
However, for sufficiently weak weight decay, 
it attains a minimum at a finite learning time 
before reaching a higher steady-state value.
This is called {\it overtraining} 
since at the later stage of learning, 
the perceptron is focusing too much 
on the specific details of the training set. 
In this case $\epsilon_g$ can be optimized by {\it early stopping}, 
i.e. terminating the learning process before it reaches the steady state. 
Similar behavior is observed in linear perceptrons \cite{krogh,bos,bos2}.

This phenomenon can be controlled by tuning the weight decay $\lambda$. 
The physical picture is that the perceptron with minimum $\epsilon_g$ 
corresponds to a point with a magnitude $|\vec J^*|$. 
When $\lambda$ is too strong, $|\vec J|$ never reaches this magnitude 
and $\epsilon_g$ saturates at a suboptimal value. 
On the other hand, when $\lambda$ is too weak, 
$|\vec J|$ grows with learning time 
and is able to pass near the optimal point during its learning history. 
Hence the weight decay $\lambda_{ot}$ for the onset of overtraining 
is closely related to the optimal weight decay $\lambda_{opt}$ 
at which the steady-state $\epsilon_g$ is minimum. 
Indeed, at $T=0$ and for all values of $\alpha$, 
$\lambda_{ot}=\lambda_{opt}=\pi/2-1$; 
the coincidence of $\lambda_{ot}$ and $\lambda_{opt}$ 
is also observed previously \cite{krogh}. 
Early stopping for $\lambda<\lambda_{ot}=\lambda_{opt}$ 
can speed up the learning process, 
but cannot outperform the optimal result at the steady state.
A recent empirical observation confirms that 
a careful control of the weight decay may be better than 
early stopping in optimizing generalization \cite{hansen}.

At nonzero temperatures, we find the new result that
$\lambda_{ot}$ and $\lambda_{opt}$ may become different.
While various scenarios are possible, 
here we only mention the case of sufficiently large $\alpha$. 
As shown in the inset of Fig. 1, 
$\lambda_{opt}$ lies inside the region of overtraining, 
implying that even the best steady-state $\epsilon_g$ 
is outperformed by some point during its own learning history. 
This means the optimal $\epsilon_g$ can only be attained 
by tuning {\it both} the weight decay and learning time. 
However, at least in the present case, 
computational results show that the improvement is marginal.

\begin{figure} [hbt]
\centering
\centerline{\psfig{figure=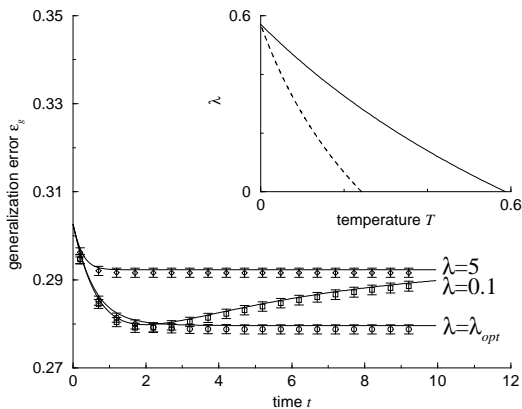,height=5.5cm}}
\caption{
The evolution of the generalization error 
at $\alpha=0.8$ and $T=0$ for different weight decay strengths $\lambda$. 
Theory: solid line, simulation: symbols.
Inset: The temperature dependence
of the optimal weight decay $\lambda_{opt}$ (dashed) and
the onset of overtraining $\lambda_{ot}$ (solid) at $\alpha=5$.
}
\end{figure}

{\it 2) Overtraining of $\epsilon_{test}$:} 
This is best understood by considering the effects 
of tuning the input noise from zero, 
when $\epsilon_{test}$ starts to increase from $\epsilon_t$. 
At the steady state $\epsilon_t$ 
is optimized by $\lambda=0$ for $\alpha<1$, 
and by a relatively small $\lambda>0$ for $\alpha>1$. 
This means that $\epsilon_{test}$ is optimized 
with no or only little concern about the magnitude of $J^2$. 
However, when input noise is introduced, 
it adds a Gaussian noise of variance $\Delta^2J^2$ 
to the activation distribution. 
The optimization of $\epsilon_{test}$ now involves 
minimizing the error of the training set 
without using an excessively large $J^2$. 
Thus the role of weight decay becomes important. 
Indeed, at $T=0$, $\lambda_{opt}=\alpha\Delta^2$ for random examples, 
whereas $\lambda_{opt}\propto\Delta^2$ approximately 
for teacher-generated examples. 
This illustrates how the environment in anticipated applications, 
i.e. the level of input noise, 
affects the optimal choice of perceptron parameters.

Analogous to the dynamics of $\epsilon_g$, 
overtraining can occur when a sufficiently weak $\lambda$ 
allows $\vec J$ to pass near the optimal point 
during its learning history. 
Indeed, at $T=0$ the onset of overtraining is given by 
$\lambda_{ot}=\lambda_{opt}$ for random examples, 
whereas $\lambda_{ot}\approx\lambda_{opt}$ for teacher-generated examples.
At nonzero temperatures, $\lambda_{ot}$ and $\lambda_{opt}$ 
become increasingly distinct, 
and for sufficiently large $\alpha$, 
$\lambda_{opt}<\lambda_{ot}$ as shown in the inset of Fig. 2, 
so that the optimal $\epsilon_{test}$ can only be attained 
by tuning {\it both} the weight decay and learning time. 

{\it 3) Average dynamics:} 
When learning has reached steady-state,
the dynamical variables fluctuates about their temporal averages
because of thermal noises.
If we consider a perceptron constructed 
using the thermally averaged weights $\langle J_j\rangle_{th}$, 
we can then prove that it is equivalent to the perceptron 
obtained at $T=0$. 
This equivalence implies that for perceptrons with thermal noises, 
the training and generalization errors can be reduced 
by temporal averaging down to those at $T=0$.

We can further compute the performance improvement 
as a function of the duration $\tau$ of the monitoring period 
for thermal averaging, 
as confirmed by simulations in Fig. 2. 
Note that the Green's function is a superposition of relaxation modes 
$\exp(-kt)$ whose rate $k$ lies in the range $k_{min}\le k\le k_{max}$, 
where $k_{max}$ and $k_{min}$ 
are $\lambda+(\sqrt\alpha\pm 1)^2$ respectively.
For $\alpha<1$, there is an additional relaxation mode with rate $\lambda$, 
which describes the relaxation by weight decay 
inside the $N-p$ dimensional solution space of zero training error. 
Hence the monitoring period scales as $k_{min}^{-1}$ for $\alpha>1$, 
and $\lambda^{-1}$ for $\alpha<1$.
Note that this time scale diverges for vanishing weight decay at $\alpha<1$. 
The time scale for thermal averaging agrees with the relaxation time 
proposed for asymptotic dynamics in \cite{hertz}.

We remark that the relaxation time for steady-state dynamics 
may not be the same as the convergence time for learning 
in the transient regime. 
For example, for $\alpha<1$ and vanishing weight decay at $T=0$, 
significant reduction of $\epsilon_t$ takes place 
in a time scale independent of $\lambda$, 
since the dynamics is dominated by a growth 
of the projection onto the solution space of zero training error. 
On the other hand, the asymptotic relaxation time diverges as $\lambda^{-1}$.

\begin{figure} [hbt]
\centering
\centerline{\psfig{figure=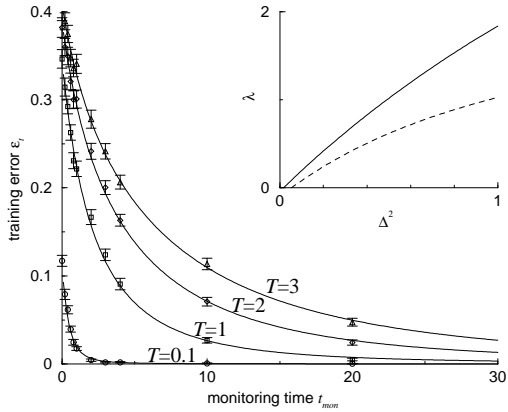,height=5.5cm}}
\caption{
The training error at $\alpha=0.1$ and $\lambda=5$ 
of the thermally averaged perceptron for random examples
versus the duration $\tau$ of the monitoring period
for thermal averaging.
Inset: The lines of the optimal weight decay $\lambda_{opt}$ (dashed)
and the onset of overtraining $\lambda_{ot}$ (solid)
of the test error for teacher-generated examples at $\alpha=3$ and $T=0.3$.
}
\end{figure}

{\it 4) Dynamics of the bias:} 
For biased perceptrons, 
$\theta(t)$ approaches the steady-state value ${\rm erf}(\phi/\sqrt 2)$. 
(The failure of the student to learn the teacher bias 
is due to the inadequacy of Adaline rule, 
and will be absent in other learning rules such as back-propagation.) 

The absence of bias decay modifies the dynamics of learning. 
For $\lambda<\sqrt\alpha-1$, $\theta(t)$ consists of relaxation modes 
with rates $k$ in the range $k_{min}\le k\le k_{max}$, 
as in the evolution of the weights. 
Hence the weights and the bias learn at the same rate, 
and convergence is limited by the rate $k_{min}$. 
However, for $\lambda>\sqrt\alpha-1$, 
$\theta(t)$ has an additional relaxation mode 
with rate $\tilde\lambda=\alpha\lambda/(1+\lambda)$. 
Since $\tilde\lambda<k_{min}$, 
the bias learns slower than the weights, 
and convergence is limited by the rate $\tilde\lambda$, 
as illustrated in Fig. 3 
which compares the evolution of the weight overlap $R(t)$ and $\theta(t)$.
If faster convergence is desired, 
the learning rate of the bias has to be increased.

\begin{figure}[hbt]
\centering
\centerline{\psfig{figure=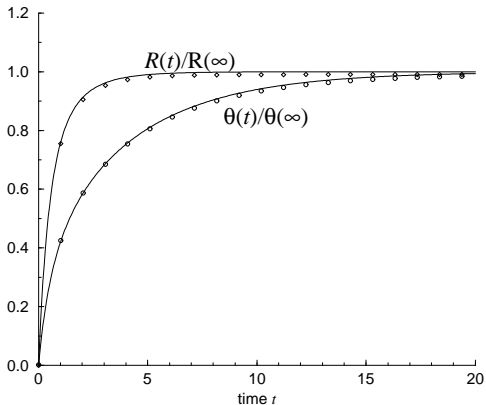,height=5.5cm}}
\caption{
The evolution of the teacher-student weight overlap $R(t)$
and the bias $\theta(t)$ at $\alpha=0.8$, $\lambda=0.4$ and $T=0$.
}
\end{figure}


In summary, 
we have introduced a general framework
for modeling the dynamics of learning based on the cavity method, 
which is much more versatile than existing theories.
It allows us to reach useful conclusions 
about overtraining and early stopping, 
input noise and temperature effects, 
transient and average dynamics, 
and the convergence of bias and weights.
We consider the present work as only the beginning of a new area of study.
Many interesting and challenging issues remain to be explored.
For example, it is interesting to generalize the method 
to dynamics with discrete learning steps of finite learning rates. 
Furthermore, the theory can be extended to multilayer networks.

This work was supported by the Research Grant Council of Hong Kong 
(HKUST6130/97P).


\end{document}